\documentclass[conference]{IEEEtran}

\usepackage{graphicx}

\newtheorem{theorem}{Theorem}[section]

\newenvironment{proof}[1][Proof]{\begin{trivlist}
\item[\hskip \labelsep {\bfseries #1}]}{\end{trivlist}}

\newcommand{\qed}{\nobreak \ifvmode \relax \else
      \ifdim\lastskip<1.5em \hskip-\lastskip
      \hskip1.5em plus0em minus0.5em \fi \nobreak
      \vrule height0.75em width0.5em depth0.25em\fi}

\begin{document}

\title{''Oh Tanenbaum, oh Tanenbaum...'': \\Technical Foundations of Xmas 4.0 Research}
\author{\IEEEauthorblockN{P. Reichl}
\IEEEauthorblockA{Research Group COSY\\Faculty for Computer Science \\University of Vienna, Austria\\
Email: peter.reichl@univie.ac.at}
\and
\IEEEauthorblockN{S. Claus}
\IEEEauthorblockA{Research Group WST\\Faculty for Computer Science \\University of Vienna, Austria\\
Email: santa.claus@cs.univie.ac.at}}

\maketitle

\begin{abstract}
Andrew Tanenbaum and his textbooks -- e.g. on Operating Systems, Computer Networks, Structured Computer Organization and Distributed Systems, to name but a few -- have had a tremendous impact on generations of computer science students (and teachers at the same time). Given this, it is striking to observe that this comprehensive body of work apparently does not provide a single line on a research topic that seems to be intimately related with his name (at least in German), i.e. Xmas Research (XR). Hence, the goal of this paper is to fill this gap and provide insight into a number of paradigmatic XR research questions, for instance: Can we today still count on Santa Claus? Or at least on Xmas trees? And does this depend  on basic tree structures, or can we rather find solutions on the level of programming languages? By addressing such basic open issues, we aim at providing a solid technical foundation for future steps  towards the imminent evolution towards Xmas 4.0. 

\end{abstract}

\IEEEpeerreviewmaketitle

\section{Introduction}

As pointed out in previous work (Reichl \& Claus 2016), the rapid evolution of the {\it Internet of Things} together with the new paradigm of {\it Industry 4.0} is considered a key milestone within the current Digital Change and will have significant impact on almost all areas of everyday life, including Xmas. Unfortunately, this application field has been largely neglected by academia and industry, and, against all odds, is not even mentioned in the comprehensive work of Andrew Tanenbaum (Tanenbaum et al. 2012, 2016 et al.), whose name intuitively is closely linked to this type of questions (at least in German). Therefore, in this paper, we try to close this gap and present several key achievements in computing technology with high relevance for Santa Claus (S.C.) operation.

At the same time, some of the puzzling mysteries around S.C. himself will be resolved as well. For instance, conclusive evidence is provided that S.C.'s characteristic oral utterances, which are widely observed and usually transcribed as ``Ho! Ho! Ho!'' without assigning them any further significance, indeed may be interpreted as an esoteric programming language of its own and in fact provide a new variant to the well-known {\it Brainf*ck} ({\it BF}) language.

The remainder of the paper is structured as follows: Starting from the computing needs of S.C. which go well beyond the addition-based mechanical calculators that have been predominant from the early ages of computer science, two different state of the art devices are presented, i.e. the {\it Xmas-Tree Calculator} and the {\it Educated Santa Claus}. As second key contribution, the S.C. specific programming language {\it Hohoho!} is discussed, and {\it Simple Hoho} is introduced as a simplified version. The paper ends with some concluding remarks and an outlook on future work.

\section{Hardware Aspects}

Concerning hardware solutions for S.C., electricity supply clearly becomes the dominating issue: for obvious reasons, only mobile devices may be considered while, on the other hand, the foreseeable heavy power consumption due to the algorithmic complexity of the problems to be solved, together with the lack of appropriate batteries available in the market today,  point to mechanical calculating devices as platform of choice, cf. Williams (1997).

\begin{figure}[ht!]
	\centering
	\includegraphics[width=6cm]{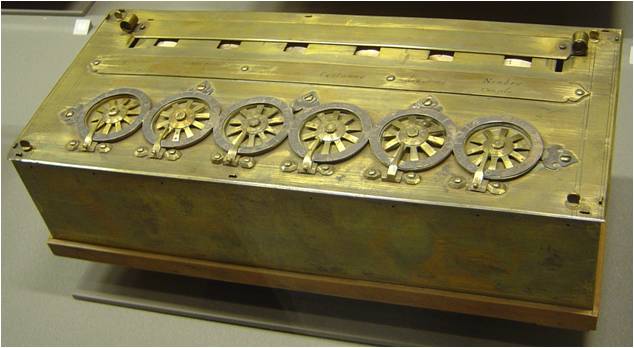}
	\caption{Pascaline, due to eminent French mathematician Blaise Pascal (1652)}
	\label{fig:Pascaline}
\end{figure}

However, determining the optimal hardware for S.C. is a non-trivial task: for instance, the {\it Pascaline} depicted in Fig. \ref{fig:Pascaline} convinces both in terms of stylish design as well as functionality, while, unfortunately, only very few copies have been produced, without any spare parts available in case some repairing should be necessary.

\begin{figure}[ht!]
	\centering
	\includegraphics[width=6cm]{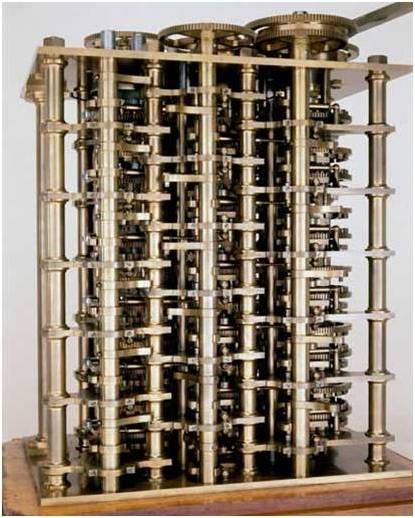}
	\caption{Analytical Engine, due to Charles Babbage (1837)}
	\label{fig:Babbage}
\end{figure}

This problem is even worse with respect to the {\it Analytical Engine} depicted in Fig. \ref{fig:Babbage}, which so far has only be realized as incomplete trial version. Despite of its remarkable precision and versality, its weight of several tons would of course pose massive statical requirements to the underlying sleigh, not to speak of the cost for the additional reindeers required.

\begin{figure}[ht!]
	\centering
	\includegraphics[width=6cm]{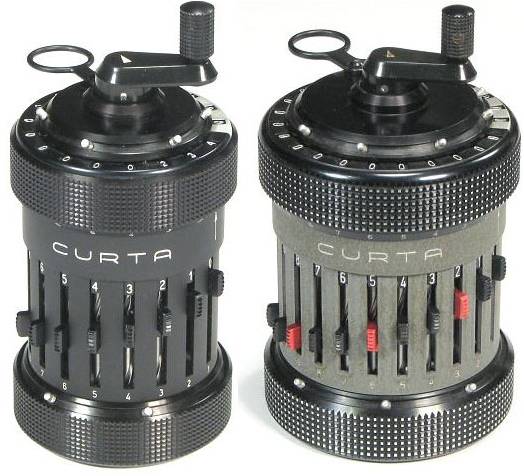}
	\caption{Curta I and Curta II, due to Curt Herzstark (1947)}
	\label{fig:Curta}
	\end{figure}

Finally, as an ``Austrian solution'', the {\it Curta} (see Fig. \ref{fig:Curta}) has become the first digital pocket calculator and a.o. for many years has been preferred by rallye drivers because of its remarkable usability and robustness also in high-speed scenarios. However, also here several problems deserve attention: first of all, S.C. would fail already with opening the associated container, as it revolves in counter-clockwise direction only. Apart from that, the {\it Curta} as well as all other mechanical devices discussed so far relies on addition as basic operational mode, whereas the typical requirements posed by S.C.-related processes clearly require a multiplication-based solution.

Summarizing, while using electronic devices is not an option for S.C. due to lack of sufficient energy supply, not even state of the art mechanical devices seem to meet the mentioned specific requirements. Therefore, we have to resort to truly disruptive technologies, which will be introduced now: (a) the Xmas Tree Calculator and (b) the Educated Santa Claus platform.

\subsection{Xmas Tree Calculator}

Using Xmas trees for calculation purposes brings along several advantages: (1) they are widely available at the appropriate seasons, avoiding any just-in-time supply issues; (2) they are extremely easy to use and at the same time incredibly robust; and (3) while operating in a fully mechanical way, they still provide sufficient enlightment based on the award-winning {\it CANDLE light} technology.  

	\begin{figure}[ht!]
	\centering
	\includegraphics[width=6.5cm]{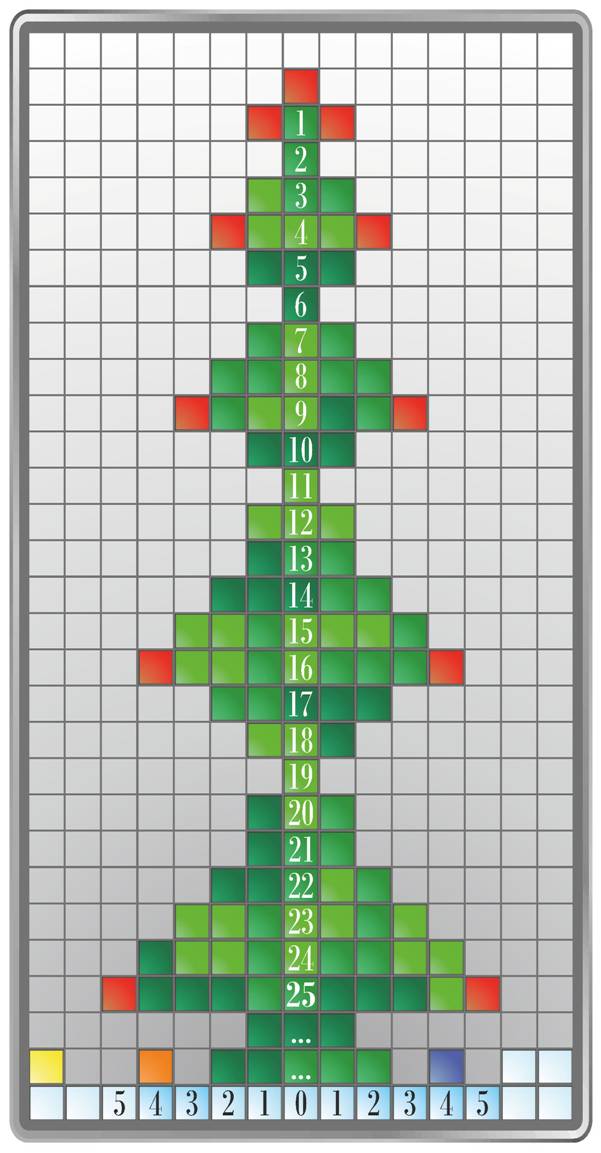}
	\caption{The ``Tetris Tree'' -- an early implementation of a Xmas Tree Calculator}
	\label{fig:Tetris}
	\end{figure}

Fig. \ref{fig:Tetris} illustrates the underlying calculation procedure which is surprisingly simple: in order to multiply two numbers from the x-axis, the corresponding red christmas balls have to be connected by a straight line (e.g. using appropriate {\it Lametta}). Then, the resulting product can be obtained from the intersection point of the {\it Lametta} with the trunk of the tree. 
For instance, in order to calculate 5 x 3, connect the red balls for ``5'' (left hand side) and ``3'' (right hand side), and the result ``15'' emerges from the intersection with the trunk. \\

\begin{theorem} The straight line connecting any two points on the convex hull of the {\it Xmas Tree Calculator} associated to values $a$ and $b$ on the x-axis, resp., crosses the trunk at $a \cdot b$.
\end{theorem}

\begin{figure}[ht!]
	\centering
	\includegraphics[width=5.5cm]{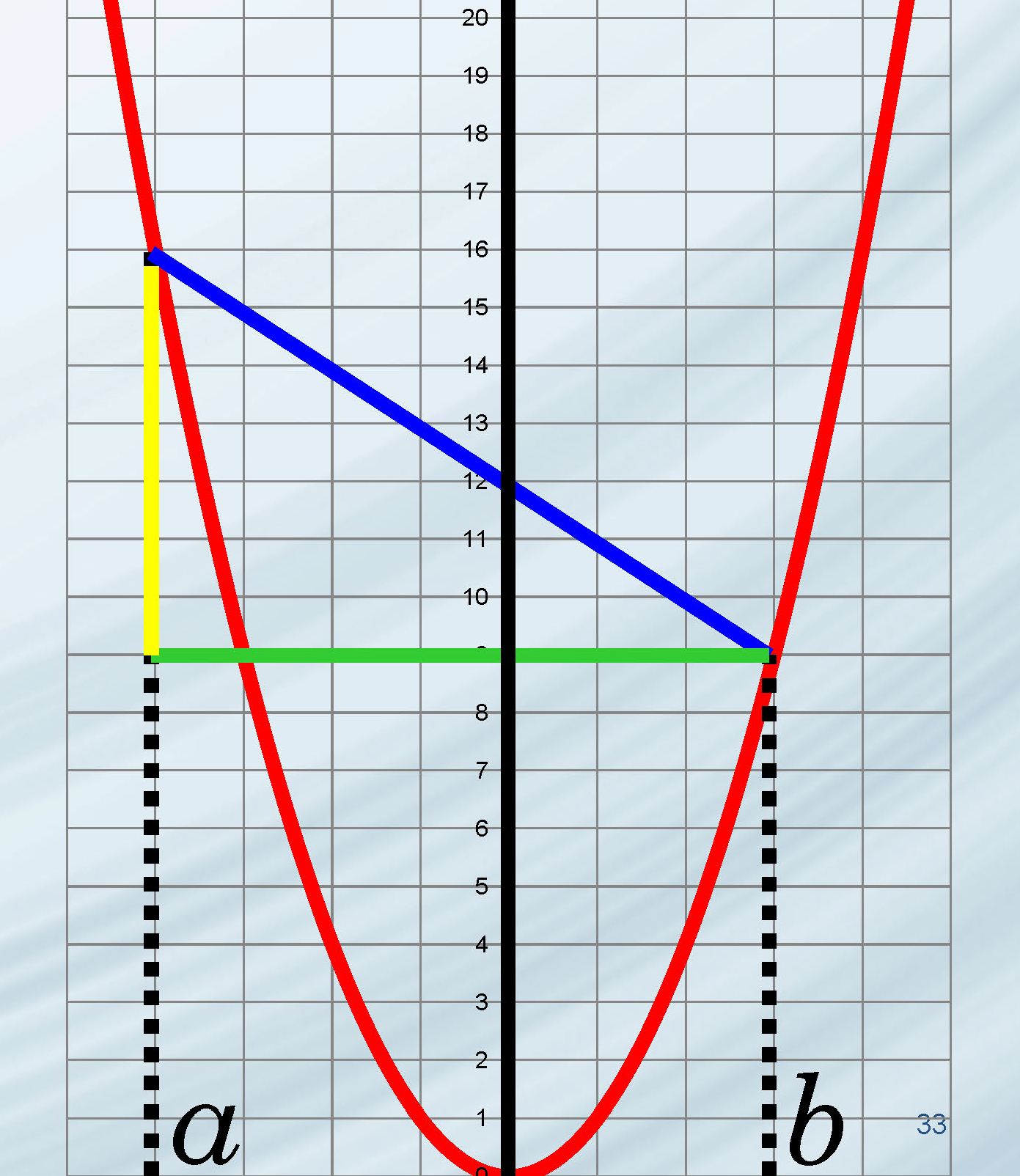}
	\caption{Geometry of the Xmas Tree Calculator}
	\label{fig:xt}
	\end{figure}

\begin{proof}:
Fig. \ref{fig:xt} presents the geometry of the {\it Xmas Tree Calculator}. Observe that the tree's convex hull (red) essentially is an inverted parabola. Hence, we can easily determine the length of the green line as $a+b$ and the length of the yellow line as $a^2-b^2$. Thus, the slope of the (descending) blue line equals

\begin{equation}
s = -\frac{a^2-b^2}{a+b}
\end{equation}
and therefore the blue line crosses the y-axis at

\begin{equation}
y = a^2-a\cdot s = a^2-a\cdot\frac{(a+b)(a-b)}{a+b} = a^2-a\cdot(a-b)=a\cdot b
\end{equation}
which refers precisely to the desired product. \qed
\end{proof}

\begin{figure}[ht!]
	\centering
	\includegraphics[width=4.07cm]{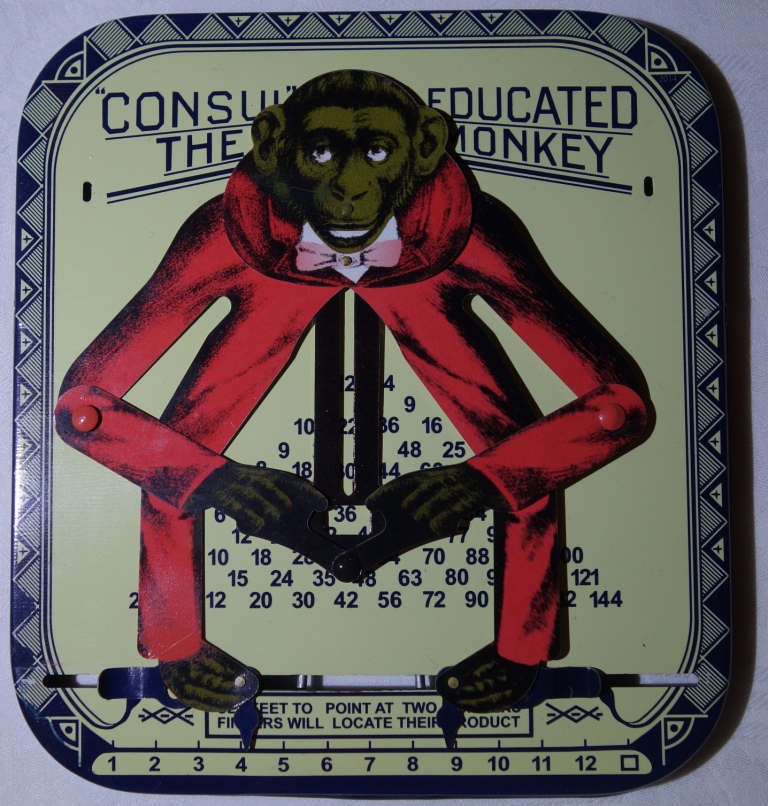}
	\includegraphics[width=4cm]{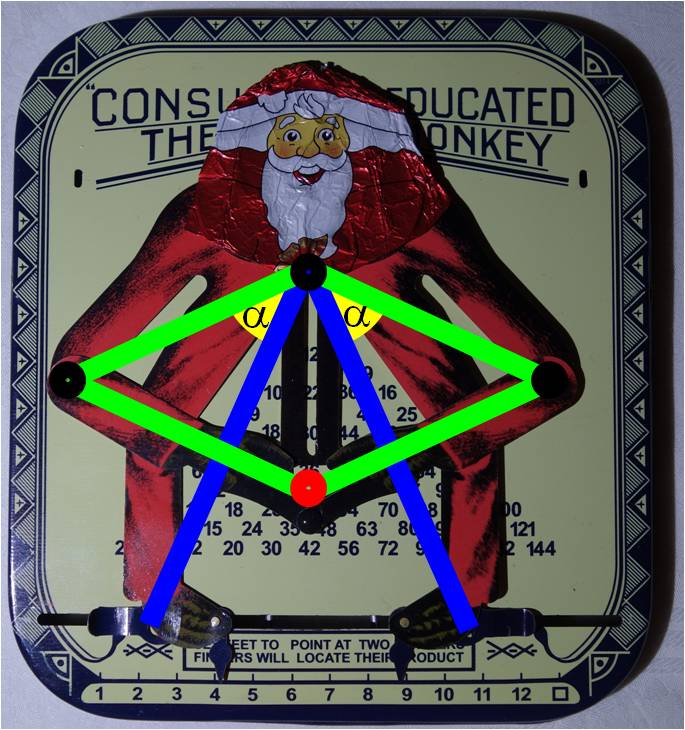}
	\caption{The Educated Monkey (left) and the Santa Claus version (right)}
	\label{fig:monkey}
	\end{figure}
	
\subsection{The Educated Santa Claus}
While the {\it Xmas Tree Calculator} provides a reliable platform for product calculation, it is still considered helpful to have an alternative option in case of availability problems. To this end, we now present our adaptation of a commercial multiplication device developed in the U.S.A. around 1915, i.e. {\it Consul -- The Educated Monkey} as depicted in Fig. \ref{fig:monkey} left.

Again, the basic idea is very straightforward: choose two numbers to be multiplied on the x-axis, put the monkey's feet upon them, and read the product from the small rectangle between the monkey's arms. \\

\begin{theorem}
Let $c$ be the length of the monkey's arms and $b$ the length of his leg, and assume $a=c/2$. For $ f =b/a=\sqrt2$, the $(x,y)$ curve of the Educated Monkey's result indicator (rectangle) is a straight line through the origin with slope 1.
\end{theorem}
\begin{figure}[ht!]
	\centering
	\includegraphics[width=6cm]{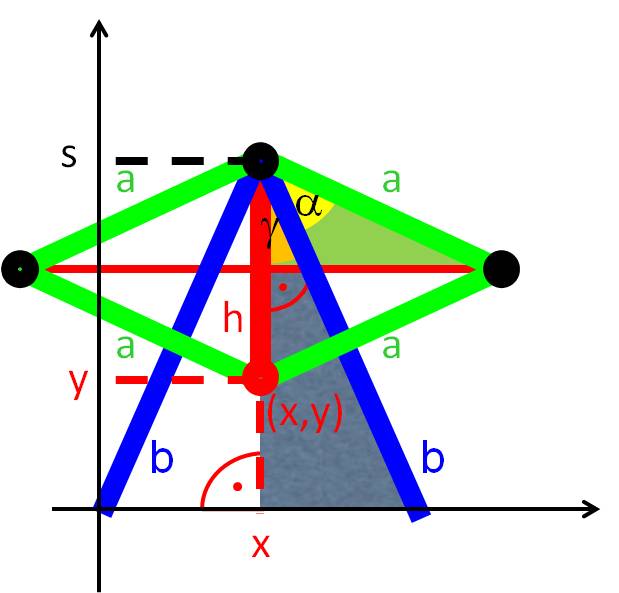}
	\caption{Geometry of the Educated Santa Claus}
	\label{fig:monkey2}
	\end{figure}
	
\begin{proof}:
Fig. \ref{fig:monkey2} depicts the geometry of the Educated Monkey/Santa Claus, where the red dot refers to the ``result indicator''. Observe that angle $\alpha$ between each arm and leg is fixed, whereas all other joints are flexible, and so is $\gamma$.

By construction, $\cos \gamma = s/b$ and $\sin \gamma = x/b$. Moreover, the green triangle is rectangular, hence
\begin{equation}
h/2 = a \cdot \cos (\alpha + \gamma) = a \cdot (\cos\alpha \cos\gamma - \sin\alpha \sin\gamma)
\end{equation}
Hence, putting everything together we get
\begin{equation}
y = s-h = b\cos\gamma - 2a\cdot (\cos\alpha \cos\gamma - \sin\alpha \sin\gamma)
\end{equation}
which, after reordering, leads to
\begin{equation}
y = (b-2a\cos\alpha)\cos\gamma + 2a \sin\alpha \cdot \frac{x}{b}
\end{equation}
For the resulting line to cross the origin it is required that 
\begin{equation}
b-2a \cos\alpha = 0 \rightarrow \cos\alpha = \frac{b}{2a}
\end{equation}
and for having slope 1:
\begin{equation}
\sin\alpha = \frac{b}{2a}.
\end{equation}
Combining equations (14) -- (15) results in $\alpha = \pi/4$, hence
\begin{equation}
b : a = \sqrt2
\end{equation}
as claimed in the theorem. \qed
\end{proof}
\vspace{12pt}
	
\begin{figure}[ht!]
	\centering
	\includegraphics[width=\columnwidth]{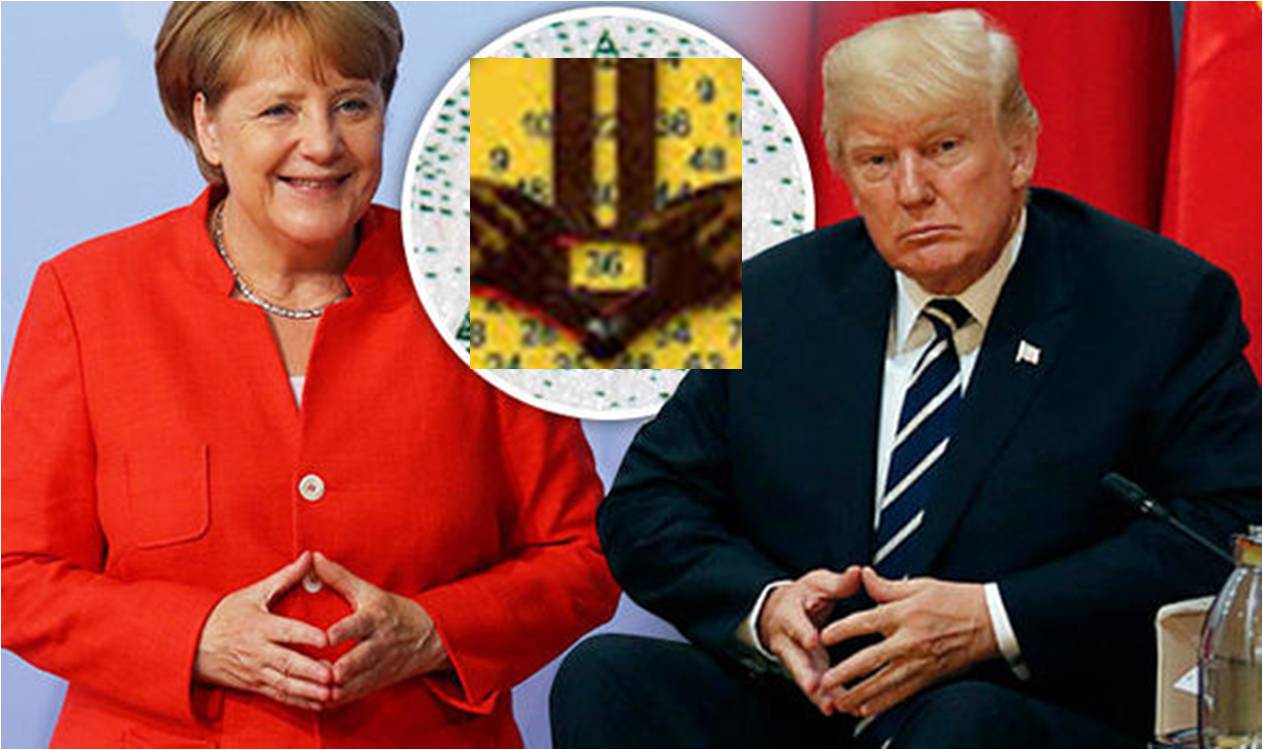}
	\caption{The Educated Monkey behaviour -- often imitated, never duplicated}
	\label{fig:Trump}
	\end{figure}

Note that, while the original form of this device has already been devised more than a century ago, and has been often imitated since (however without ever matching the original, see for instance Fig. \ref{fig:Trump}), for our purposes it needs significant redesign efforts.

\begin{figure}[ht!]
	\centering
	\includegraphics[width=5cm]{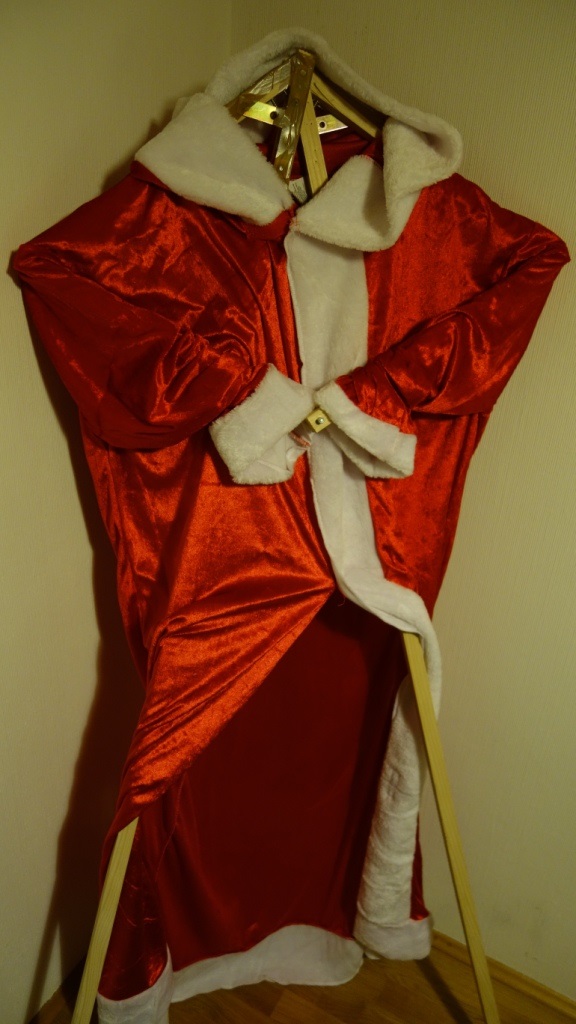}
	\caption{The Educated Santa Claus -- LoFi prototype}
	\label{fig:LoFi}
	\end{figure}

In principle, there are two implementation options: either we rely on the existing hardware platform modified using parts of a chocolate Santa Claus, which eventually provides a readily available solution and, in addition, provides certain culinary collateral advantages (see Fig. \ref{fig:monkey} right-hand side). Alternatively, Fig. \ref{fig:LoFi} depicts a LoFi prototype which has been constructed mainly for S.C. training purposes in realistic scenarios.

\section{A Programming Language for Santa Claus}

\begin{figure}[ht!]
	\centering
	\includegraphics[width=6cm]{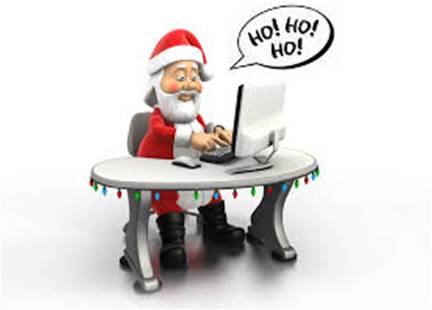}
	\caption{A Programming Language for Santa Claus?}
	\label{fig:SC_hohoho}
\end{figure}

The second contribution of this paper concerns a novel programming language especially designed for Xmas4.0 purposes. To the best of our knowledge, the question of which programming language S.C. is actually using has never been posed in a scientific context so far, while the result of a dedicated user trial is sketched in Fig. \ref{fig:SC_hohoho}.

A more in-depth analysis of the problem eventually leads to the insight that such a programming language will be used by S.C. as well as by reindeers and Xmas trees as well (cf. Section 2.2). Hence, we may refer to related work by D. Morgan-Mar and his programming language {\it Ook!} which has been designed for the usage by orang-utans. Of course, neither S.C. nor reindeers or Xmas trees belong to any of the Pongo subspecies, however, {\it Ook!} nevertheless offers excellent guidance for solving our problem.
In fact, {\it Ook!} essentially is a variant of the esoteric programming language {\it Brainf*ck} ({\it BF} for short), which represents {\it BF}'s eight different syntax elements by 2-tupels of utterances ``Ook.'', ``Ook!'' and/or ``Ook?'', resp. This is, however, an extremely inefficient way of encoding, as we have a total of $3^2 = 9$ commands at hands while {\it BF} requires only 8 different commands in total. 

\begin{figure}[ht!]
	\centering
	\includegraphics[width=\columnwidth]{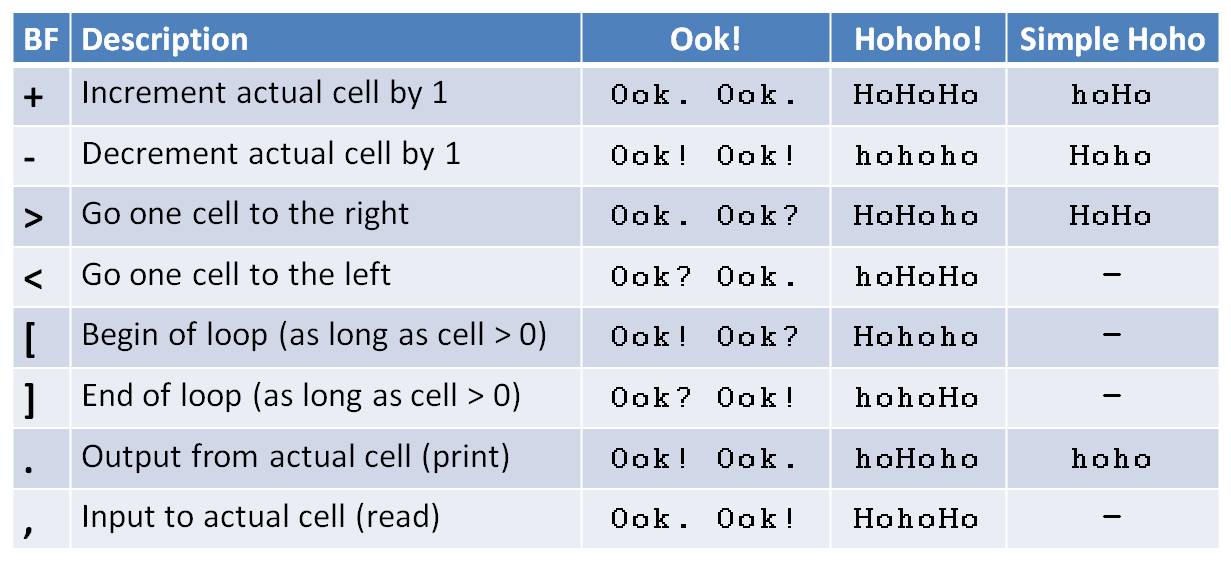}
	\caption{Syntax Elements of {\it Hohoho!} and {\it Simple Hoho}}
	\label{fig:Hoho}
\end{figure}

Subsequent extensive research work eventually leads to the conclusion that S.C. has found a surprisingly simple way to improve on this, i.e. by using 3-tupels of ``Ho'' and ``ho'', respectively, and thus employing an optimal encoding scheme fully compatible with {\it BF} as depicted in Fig. \ref{fig:Hoho}.

In order to increase readability, the following advanced notation is proposed: by definition, each {\it Hohoho!} command consists of three subsequent ``Ho's'' and/or ``ho's'', resp.; call these basic units ``atoms''. Keeping this in mind, we easily can separate any sequence of atoms, and similarly join them in an arbitrary way, as long as a {\it Hohoho!} compiler extracts just subsequent triplets of atoms and interpretes them as individual commands. Hence, we may rewrite any {\it Hohoho!} program such that an upper case ``Ho'' always is followed by an arbitrary number (larger or equal to zero) of lower case ``ho's'' plus an exclamation mark and a space character. In this notation, e.g. ``Ho! Hohoho! Hoho!'' actually refers to ``HoHoho hoHoho'' in standard {\it Hohoho!} and corresponds to the {\it BF} command sequence ``$>.$''. To further illustrate this, here is an example program in {\it Hohoho!}, printing ``Hello World!'' to the standard output:

\vspace{8pt}

{\tiny \noindent Ho! Ho! Ho! Ho! Ho! Ho! Ho! Ho! Ho! Ho! Ho! Ho! Ho! Ho! Ho! Ho! Ho! Ho! Ho! Ho! Ho! Ho! Ho! Ho! Ho! Ho! Ho! Ho! Ho! Ho! Hohoho! Ho! Hoho! Ho! Ho! Ho! Ho! Ho! Ho! Ho! Ho! Ho! Ho! Ho! Ho! Ho! Ho! Ho! Ho! Ho! Ho! Ho! Ho! Ho! Ho! Hoho! Ho! Ho! Ho! Ho! Ho! Ho! Ho! Ho! Ho! Ho! Ho! Ho! Ho! Ho! Ho! Ho! Ho! Ho! Ho! Ho! Ho! Ho! Ho! Ho! Ho! Ho! Ho! Ho! Ho! Ho! Ho! Hoho! Ho! Ho! Ho! Ho! Ho! Ho! Ho! Ho! Ho! Ho! Hoho! Ho! Ho! Hohoho! Hohoho! Hohoho! Hohoho! Hohohohohoho! Ho! Ho! Hoho! Ho! Ho! Ho! Ho! Ho! Hoho! Hoho! Ho! Hoho! Ho! Ho! Hoho! Hoho! Ho! Ho! Ho! Ho! Ho! Ho! Ho! Ho! Ho! Ho! Ho! Ho! Ho! Ho! Ho! Ho! Ho! Ho! Ho! Ho! Hoho! Hohoho! Hoho! Ho! Ho! Ho! Ho! Ho! Ho! Ho! Ho! Hoho! Hoho! Ho! Hoho! Ho! Ho! Ho! Ho! Ho! Hoho! Hohohoho! Hohoho! Ho! Ho! Ho! Ho! Ho! Ho! Ho! Ho! Ho! Ho! Ho! Ho! Ho! Ho! Ho! Ho! Ho! Ho! Ho! Ho! Ho! Ho! Ho! Ho! Ho! Ho! Ho! Ho! Ho! Ho! Ho! Ho! Ho! Ho! Ho! Ho! Ho! Ho! Ho! Ho! Ho! Ho! Ho! Ho! Ho! Hoho! Hoho! Ho! Hohoho! Hoho! Ho! Ho! Ho! Ho! Ho! Ho! Ho! Ho! Hoho! Hohohohohohohohohohohohohohohohohohohohoho! Hohohohohohohohohohohohohohohohohohohohohohohohohohoho! Hoho! Ho! Hoho! Ho! Ho! Hoho! Hoho!} \\

While {\it Hohoho!}, like {\it BF} and {\it Ook!}, is a Turing-complete language, in fact S.C. does not employ many of the basic commands, for the following reasons:
\begin{itemize}
\item Loops are much more useful for Xmas carols and hence avoided by S.C.;
\item Going backward with a sleigh is extremely difficult and hence is carefully avoided by S.C. as well;
\item S.C.'s occupation implies output (presents $\rightarrow$ children) rather than input operations.
\end{itemize}

As a consequence, the number of actual commands can be reduced to four, leading to {\it Simple Hoho} as an even more efficient programming language, see Fig. \ref{fig:Hoho}. 
Hence, the above program example, now in {\it Simple Hoho} with analogous advanced 2-tupel notation, reads as follows:
\vspace{8pt}

{\tiny \noindent Ho! Hoho! Hoho! Hoho! Hoho! Hoho! Hoho! Hoho! Hoho! Hoho! Hoho! Hoho! Hoho! Hoho! Hoho! Hoho! Hoho! Hoho! Hoho! Hoho! Hoho! Hoho! Hoho! Hoho! Hoho! Hoho! Hoho! Hoho! Hoho! Hoho! Hoho! Hoho! Hoho! Hoho! Hoho! Hoho! Hoho! Hoho! Hoho! Hoho! Hoho! Hoho! Hoho! Hoho! Hoho! Hoho! Hoho! Hoho! Hoho! Hoho! Hoho! Hoho! Hoho! Hoho! Hoho! Hoho! Hoho! Hoho! Hoho! Hoho! Hoho! Hoho! Hoho! Hoho! Hoho! Hoho! Hoho! Hoho! Hoho! Hoho! Hoho! Hoho! Hoho! Hohohoho! Hoho! Hoho! Hoho! Hoho! Hoho! Hoho! Hoho! Hoho! Hoho! Hoho! Hoho! Hoho! Hoho! Hoho! Hoho! Hoho! Hoho! Hoho! Hoho! Hoho! Hoho! Hoho! Hoho! Hoho! Hoho! Hoho! Hoho! Hoho! Hohohoho! Hoho! Hoho! Hoho! Hoho! Hoho! Hoho! Hohohohohoho! Hoho! Hoho! Hohoho! Hoho! Hoho! Hoho! Hoho! Hoho! Hoho! Hoho! Hoho! Hoho! Hoho! Hoho! Hoho! Hoho! Hoho! Hoho! Hoho! Hoho! Hoho! Hoho! Hoho! Hoho! Hoho! Hoho! Hoho! Hoho! Hoho! Hoho! Hoho! Hoho! Hoho! Hoho! Hoho! Hoho! Hoho! Hoho! Hoho! Hoho! Hoho! Hoho! Hoho! Hoho! Hoho! Hoho! Hoho! Hoho! Hoho! Hoho! Hoho! Hoho! Hoho! Hoho! Hoho! Hoho! Hoho! Hoho! Hoho! Hoho! Hoho! Hoho! Hoho! Hoho! Hoho! Hoho! Hoho! Hoho! Hoho! Hoho! Hoho! Hoho! Hoho! Hoho! Hoho! Hoho! Hoho! Hoho! Hoho! Hoho! Hoho! Hohohohoho! Hoho! Hoho! Hoho! Hoho! Hoho! Hoho! Hoho! Hoho! Hoho! Hoho! Hoho! Hoho! Hoho! Hoho! Hoho! Hoho! Hoho! Hoho! Hoho! Hoho! Hoho! Hoho! Hoho! Hoho! Hoho! Hoho! Hoho! Hoho! Hoho! Hoho! Hoho! Hoho! Hoho! Hoho! Hoho! Hoho! Hoho! Hoho! Hoho! Hoho! Hoho! Hoho! Hoho! Hoho! Hoho! Hoho! Hoho! Hoho! Hoho! Hoho! Hoho! Hoho! Hoho! Hoho! Hohohoho! Hoho! Hoho! Hoho! Hoho! Hoho! Hoho! Hoho! Hoho! Hoho! Hoho! Hoho! Hoho! Hoho! Hoho! Hoho! Hoho! Hoho! Hoho! Hoho! Hoho! Hoho! Hoho! Hoho! Hohohoho! Hoho! Hoho! Hohoho! Hoho! Hoho! Hoho! Hoho! Hoho! Hohohoho! Hoho! Hoho! Hoho! Hoho! Hoho! Hoho! Hoho! Hohohoho! Hoho! Hoho! Hoho! Hoho! Hoho! Hoho! Hoho! Hoho! Hoho! Hoho! Hoho! Hoho! Hoho! Hoho! Hoho! Hoho! Hoho! Hoho! Hoho! Hoho! Hoho! Hoho! Hoho! Hoho! Hoho! Hoho! Hoho! Hoho! Hoho! Hoho! Hoho! Hoho! Hoho! Hoho! Hoho! Hoho! Hoho! Hoho! Hoho! Hoho! Hoho! Hoho! Hoho! Hoho! Hoho! Hoho! Hoho! Hoho! Hoho! Hoho! Hoho! Hoho! Hoho! Hoho! Hoho! Hoho! Hoho! Hoho! Hoho! Hoho! Hoho! Hoho! Hoho! Hoho! Hoho! Hoho! Hohohoho!}\\

\begin{figure}[ht!]
	\centering
	\includegraphics[width=3.8cm]{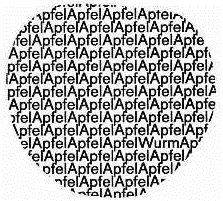}
		\includegraphics[width=4.2cm]{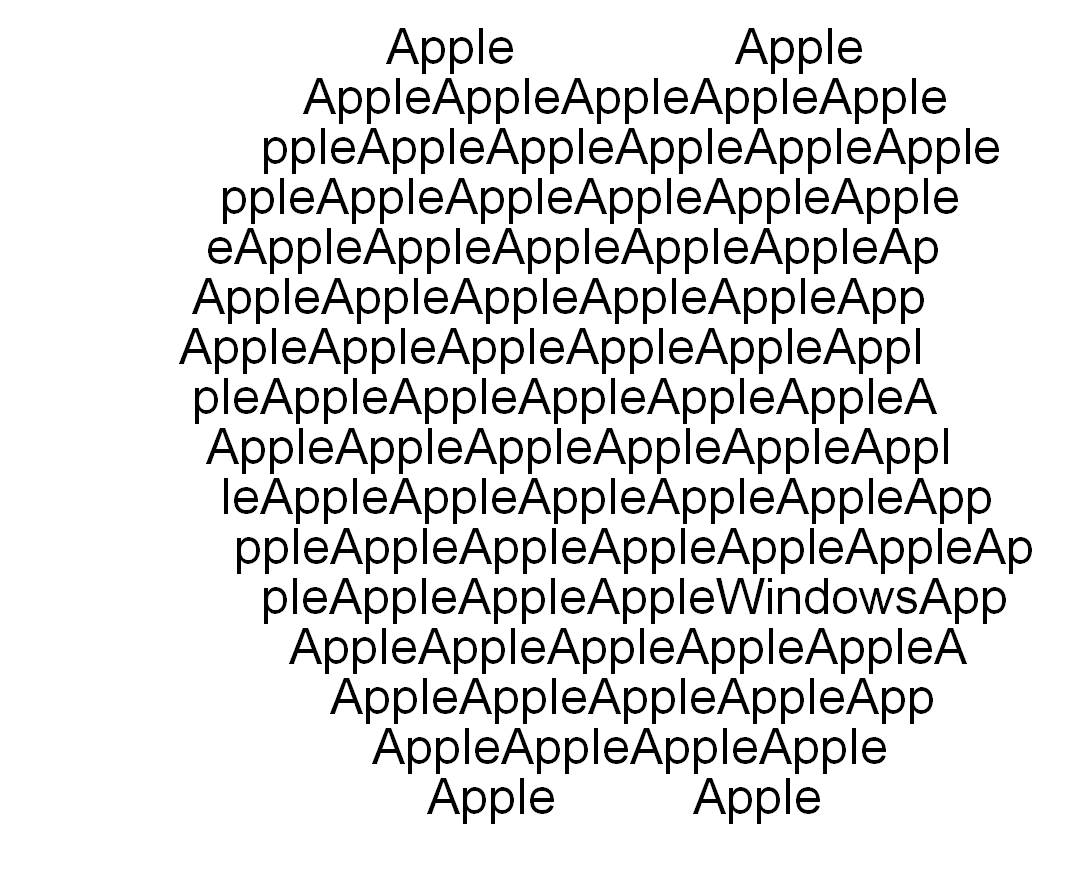}
	\caption{``apfel mit wurm'' by Doehl (1965) and translation by Reichl (2017).}
	\label{fig:Apfel}
\end{figure}

From this example, it is interesting to observe that programming in {\it Simple Hoho} not only may be considered a primary example for the new paradigm of {\it loop-free programming}, but at the same time leads to results that are also visually appealing, and thus does not only constitute a (somewhat unexpected) link to Br\"unnhilde's {\it Hojotoho} cries in Richard Wagner's ``Ring'' cycle (proving once more that everything that can be said has already been said in opera), but also to so-called {\it concrete poetry}. For further evidence we refer to Fig. \ref{fig:Apfel} which depicts one of the most representative examples of this literary genre due to Reinhard Doehl (together with a recent translation into English by the corresponding author).

\section{Conclusions}

In order to fill an apparent gap in the comprehensive work of A. Tanenbaum, this paper addresses several basic issues concerning the technical foundations of Xmas 4.0 research. As key contributions it is demonstrated how to do multiplications between arbitrary numbers with the help of Xmas trees and using a Santa Claus-specific variant of the well-established {\it Educated Monkey} mechanism. Moreover, the novel esoteric programming language {\it Hohoho!} is described, as well as the simplified version {\it Simple Hoho}. Current and future work deals mainly with implementation issues of these disruptive approaches, including e.g. a dedicated {\it Hohoho!} compiler.

\section*{Acknowledgments}
\noindent
Part of this work is supported by the H2020 symbIoTe project, which has received funding from the European Union's H2020 research and innovation programme under grant agreement 688156. This paper would not have been possible without the continuous support of the entire COSY team the authors would like to thank for. Special thanks to Roman Heger from University of Duisburg-Essen for the design of the Tetris tree.

\section*{References}

\begin{footnotesize}
\noindent
P. Reichl, S. Claus (2016): {\it Towards Xmas 4.0: Recent Advances in Santa Claus Research}. Under review, Journal of Universal Rejection, 2016. \\

\noindent
A. Tanenbaum, H. Bos (2016): {\it Modern Operating Systems}. Pearson, 2016.\\

\noindent
A. Tanenbaum, T. Austin (2012): {\it Structured Computer Organization}. Pearson, 2012.\\

\noindent
M. R. Williams (1997): {\it History of Computing Technology}. IEEE Computer Society Press, 1997.\\

\noindent
R. Doehl (1965): {\it apfel mit wurm}. In: E. Gomringer (ed.), Konkrete Poesie. Stuttgart, Reclam 1972, p.38.

\end{footnotesize}
\end{document}